\documentclass[aps,pra,onecolumn,superscriptaddress]{revtex4}
\usepackage{amsmath,amssymb,amsfonts,bbm,graphicx,times,psfrag}
\usepackage[pdftex]{color}
\usepackage{amsmath,bm}
\usepackage{mathrsfs}
\usepackage[colorlinks, linkcolor=blue, citecolor=red, urlcolor=red, breaklinks=true]{hyperref}
\usepackage{microtype}
\usepackage{epstopdf}

\usepackage{amsthm}
\usepackage{csquotes}
\usepackage{amsthm}\newcommand{\be} {\begin{equation}}
\newcommand{\ee} {\end{equation}}

\usepackage{color}
\usepackage{mathrsfs}
\newcommand{\bla}{bla\\bla\\bla\\bla\\bla}

\newcommand{\tr}{\text{tr}}

\newcommand{\I}{\mathcal{I}} 
\newcommand{\Arev}{A^{\text{rev}}}
\newcommand{\Airr}{A^{\text{irr}}} 
\newcommand{\Jrev}{J^{\text{rev}}}
\newcommand{\Jirr}{J^{\text{irr}}}

\begin{document}

\title{Coherent feedback-enhanced asymmetry of thermal process in open quantum systems: Cavity optomechanics}

\author{Hamza Harraf}
\affiliation{LPHE-Modeling and Simulation, Faculty of Sciences, Mohammed V University in Rabat, Rabat, Morocco}	
\author{Mohamed Amazioug} \thanks{m.amazioug@uiz.ac.ma}
\affiliation{LPTHE-Department of Physics, Faculty of Sciences, Ibnou Zohr University, Agadir 80000, Morocco}
\author{Rachid Ahl Laamara}
\affiliation{LPHE-Modeling and Simulation, Faculty of Sciences, Mohammed V University in Rabat, Rabat, Morocco}
\affiliation{Centre of Physics and Mathematics, CPM, Faculty of Sciences, Mohammed V University in Rabat, Rabat, Morocco}

\vspace{0.1cm}

\begin{abstract}

Entropy production is a fundamental concept in nonequilibrium thermodynamics, providing a direct measure of the irreversibility inherent in any physical process. In this work, we investigate in steady-state the enhancement of irreversibility employing coherent feedback loop. We evaluate the steady-state entropy production rate and quantum correlations by applying the quantum phase space formulation to calculate the entropy change. Our study reveals the essential contribution of coherent feedback in the thermal bath's input-noise operators, resulting in the system being driven far from thermal equilibrium. Our analysis shows that in the small-coupling limit, the entropy production rate is proportional to the quantum mutual information.
We use for application the optomechanical system of Fabry-P\'erot cavity, and show that the picks of the entropy production corresponding of the heating/cooling of movable mirror are improved. Therefore, we conclude that irreversibility and quantum correlations are not independent and must be analyzed jointly. The results demonstrate the possibility of enhancement of entropy production and pave the way for promising quantum thermal applications through coherent feedback loop.	

\end{abstract}

\maketitle

\section{Introduction}

Entropy is considered as a fundamental measure that connects seemingly different notions like disorder, information, and irreversibility, playing a pivotal role in understanding dynamical processes \cite{Sethna2011,Zemansky1968}. In the microscopic picture, the total entropy of a system in state $\rho$ writes as
\begin{equation}\label{SS}
S(\rho) = - {\rm tr} (\rho {\rm log} \rho ) \, .
\end{equation}
Any transformation that occurs within a finite-time results in the production of entropy, signifying the irreversible nature of the process. The irreversible entropy generated by a process is of paramount importance. Entropy production leads for understanding nonequilibrium systems, and minimizing it can significantly enhance the efficiency of thermal machines. However, microscopic definition of entropy production is particularly challenging. Besides, entropy production is a fundamental quantity, its direct measurement remains elusive. This presents a significant challenge, especially when applying thermodynamic considerations. Furthermore, the emergence of correlations within a quantum system is intrinsically tied to the generation of entropy. This suggests that irreversibility and the formation of correlations are interconnected and cannot be considered independently. The entropy production rate, dictated by the second law of thermodynamics, must always be positive or zero \cite{Seifert2012,Jarzynski2011}. This yields the following rate equation for entropy change
\begin{equation}\label{S}
\Delta S - \int\frac{\delta Q}{T}\ge 0 \, ,
\end{equation}
where, $T$ represents the temperature of the system, and $\delta Q$ denotes the infinitesimal amount of heat absorbed by the system. The strict inequality indicates that irreversible processes are accompanied by energy dissipation, where some energy is lost to the environment as heat~\cite{Prigogine1968}. The rate of variation of entropy for a system interacting with a thermal bath is given by
\begin{equation}\label{RateEq}
\frac{\text{d}S}{\text{d} t} =\Phi(t)+\Pi(t) \, ,
\end{equation}
in this equation, $\Pi(t)$ represents the irreversible entropy production rate, while $\Phi(t)$ represents the flux of entropy from the surrounding environment. Once the system reaches a stationary state, the irreversible entropy production rate becomes $\Pi_{\text{s}}$ and the entropy flux becomes $\Phi_{\text{s}}$. In this state, $\Pi_{\text{s}} = - \Phi_{\text{s}} > 0$. Only when both $\Pi_{\text{s}}$ and $\Phi_{\text{s}}$ are zero, the system attains thermal equilibrium.\\
  
In the classical realm, entropy production is typically studied using Onsager's theory \cite{Onsager31,Onsager53}, classical master equations \cite{Schnakenberg76}, and Fokker-Planck equations \cite{Tome2010,Tome2012}. Extending these concepts to the quantum domain involves quantum master equations \cite{Lindblad1976,Leggio2013}. Within this framework, entropy production can be quantified in terms of the mean values, variances, and irreversible quasi-probability currents in phase space. This approach has been applied to single-mode Gaussian systems coupled to a single reservoir \cite{Santos2017,Edet2024,Shahidani2024PRA} and multi-mode Gaussian systems interacting with multiple reservoirs \cite{Malouf2019,Giordano2021}. This underscores that the Wigner entropy production rate stays finite, even for a system coupled to a thermal reservoir at zero temperature \cite{Santos2018}. The thermodynamic cost associated with establishing correlations between systems has been examined \cite{Huber2015}, adopting a purely information-theoretical approach without considering their dynamical origins. The generation of correlations and production of entropy are complementary aspects in a dissipative process \cite{Santos2017}. Experimental results have shown good agreement with the theoretical predictions of this formalism \cite{Brunelli2018}. \\

Significant attention has recently been paid to coherent feedback in the fields of quantum information. Wiseman first discussed coherent feedback within the optical system \cite{WisemanPRL93,WisemanPRA93}. By using an asymmetric beam splitter to direct the input light field to the cavity, coherent feedback is achieved in a Fabry-P\'erot cavity optomechanical setup. Moreover, a portion of the output electromagnetic field is reflected by the mirror and then fed back into the cavity through an asymmetric beam splitter. Feedback control is central to a wide range of applications, including the cooling of mechanical oscillators \cite{VitaliPRL98,Pinard01,VitaliPRA02} and the entanglement of superconducting qubits \cite{Riste13}. Optomechanical systems \cite{VitaliPRL17,VitaliQST17}, trapped ions \cite{Bushev06}, cooling and trapping single atoms \cite{Koch10}, trapped nanospheres \cite{Gieseler12,Genoni15}, and entanglement generation protocols \cite{Li17,Huang19,Brunelli21,Brunelli23} and quantum state transfer \cite{amazioug20} are current platforms for coherent feedback implementation.\\

This paper investigates irreversibility in a bipartite quantum system with coherent feedback. A simplified model system consisting of two linearly interacting quantum oscillators, each dissipating into a local thermal bath, is studied. This system, despite its simplicity, captures many essential features of quantum-controlled experiments. The expression for the steady-state rate of irreversible entropy production is utilized \cite{Santos2017}. This expression links irreversibility to either the number of excitations in the two modes (relative to their equilibrium values) or the correlations between their motion. This clarifies how irreversibility emerges. We prove the enhancement of non-equilibrium between the two oscillators via coherent feedback. We compare the behavior of entropy production with total correlations, which are quantified by mutual information \cite{Vedral2001,Vedral2012}. We observe that entropy production exhibits a monotonic relationship with mutual information. We find a simple proportionality between entropy production and mutual information, specifically when the coupling between the systems is weak compared to their natural frequencies. We utilize our results for the quantification of irreversibility in an optomechanical system \cite{Aspelmeyer2014}. Besides, system characteristics influence the behavior of entropy production in various regimes, such as cavity-mediated cooling and amplification of the mechanical resonator.\\

The remainder of this paper is outlined as follows: In Section II, we suggest the model and quantum Langevin equations of motion of coupled quantum oscillators. In Section III, we discuss the covariance matrix of the system, the stationary entropy production rate, and quantum correlation quantified by mutual information and negativity logarithmic. In Section IV, is devoted to discussion of the results. In Section V, we consider an optomechanical system as an application of the system under consideration. The conclusion is presented in Section VI.

\section{The Model OF COUPLED QUANTUM OSCILLATORS}\label{Sec:Model}

\begin{figure}
\centering
 \includegraphics[width=16cm,height=5cm,angle=0]{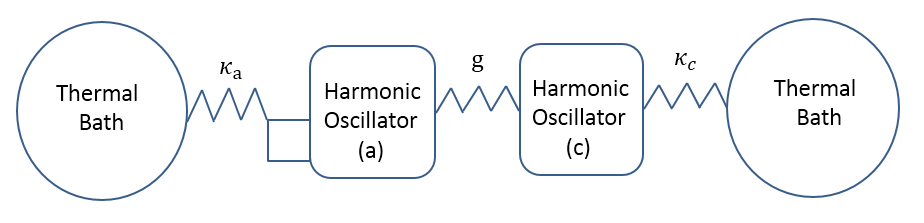}
 \caption{Schematic of the system.}
 \label{fig0} 
\end{figure} 
The system under consideration comprises two coupled harmonic quantum oscillators, each dissipating into its own local thermal bath, as depicted in Fig. \ref{fig0}. The Hamiltonian of the interacting system is described by
\begin{eqnarray}\label{H1}
\mathcal{H}= \Omega_{\rm fb} \hat{\rm a}^\dagger \hat{\rm a} + \Omega_c \hat{c}^\dagger \hat{c} - {\rm g}(\hat{\rm a} + \hat{\rm a}^\dagger)(\hat{c} + \hat{c}^\dagger),
\end{eqnarray}
where the two harmonic oscillators have annihilation (creation) operators $\hat{\rm a} (\hat{\rm a}^\dagger)$ and $\hat{c} (\hat{c}^\dagger)$, and frequencies $\Omega_{\rm fb}$ and $\Omega_{\rm c}$, respectively. Equation (\ref{H1}) describes the system's Hamiltonian, where the first two terms represent the free energies of the harmonic oscillators and the last term denotes their coupling with coupling constant ${\rm g}$. Each of the two harmonic oscillators dissipates into its local thermal bath (bath 1 at rate $\kappa_{\rm a}$, bath 2 at rate $\kappa_{\rm c}$). Assuming the baths are mutually independent, the system is also subjected to additional quantum noise, described by the input operators ${\rm a}^{in}$; $c^{in}$ satisfying Markovian correlation functions  
\begin{subequations}
 \begin{eqnarray}\label{correlation-th}
\langle \hat{\rm a}_{\rm fb}^{\rm in}(t) \, \hat{\rm a}_{\rm fb}^{\rm in \dag}(t')\rangle &=& \xi^2 |1-\tau e^{ {\rm i} \theta}|^2 \left(N_{\rm a}+1\right) \,\delta(t{-}t'),\\
\langle \hat{\rm a}_{\rm fb}^{\rm in \dag}(t) \, \hat{\rm a}_{\rm fb}^{\rm in}(t')\rangle \,\,&{=}&\,\, \xi^2 |1-\tau e^{ {\rm i} \theta}|^2 N_{\rm a} \, \delta(t{-}t'),\\
 \langle \hat{c}^{in}(t)\hat{c}^{in}(t')\rangle &=& \langle \hat{c}_{in}^{\dagger}(t)\hat{c}^{\dagger}_{in}(t')\rangle=0,\\
\langle \hat{c}^{in}(t) \hat{c}^{in \dagger}(t')\rangle &=&(N_c+1)\delta(t-t'),
\end{eqnarray}
\end{subequations}
where the thermal occupation number associated with reservoir $i$ (at temperature $T_i$, for $i={\rm a}, c$) is given by
\begin{equation}
N_{\rm a} = (e^{\hbar \Omega_{\rm a}/k_B T_{\rm a}}-1)^{-1},~~N_{c}=(e^{\hbar \Omega_c/k_B T_{c}}-1)^{-1}, 
\end{equation}
where $k_B$ is the Boltzmann constant. The beam splitter has a real amplitude transmission parameter $\xi$ and a real amplitude reflectivity parameter $\tau$. These real, positive quantities satisfy $\xi^2 + \tau^2 = 1$. Incorporating dissipation from system-bath couplings and their corresponding noises, the quantum Langevin equations of motion are given by
\begin{eqnarray}\label{La}
\dot{\hat{\rm a}}&=&-(i \Omega_{\rm fb} + \kappa_{\rm fb}) \hat{\rm a} + i {\rm g} (\hat{c}+\hat{c}^\dagger) + \sqrt{2\kappa} {\rm a}_{\rm fb}^{in}, \nonumber\\
 \dot{\hat{c}}&=&-(i \Omega_c + \kappa_c)\hat{c} +i {\rm g} (\hat{\rm a} + \hat{\rm a}^\dagger) +\sqrt{2\gamma} c^{in},\nonumber\\
\end{eqnarray}
the effective cavity decay rate and detuning are given by $\kappa_{\rm fb}=\kappa_{\rm a}(1-2\tau\cos\theta)$ and $\Omega_{\rm fb}=\Omega_{\rm a}-2\kappa_{\rm a}\tau\sin\theta$, respectively. In these expressions, $\theta$ represents the phase shift generated by the reflectivity of the output field on the mirrors. The standard input-output relation relating the output field $\hat{\rm a}^{\rm out}$ and the cavity field $\hat{\rm a}$ is $\hat{\rm a}^{\rm out} = \sqrt{2\kappa_{\rm a} } \hat{\rm a} - {\rm a}^{\rm in}$ \cite{DFWalls1998}. To describe the input field induced via the coherent feedback, we introduce the effective input noise operator $\hat{\rm a}^{\rm in}_{\rm fb}$, given by $\hat{\rm a}^{\rm in}_{\rm fb} =\tau e^{{\rm i}\theta} \hat{\rm a}^{\rm in} + \tau \hat{\rm a}^{\rm in}$.

\section{ENTROPY PRODUCTION RATE and QUANTUM MUTUAL INFORMATION}

We define the EPR-type quadrature operators for the two modes as $\hat{X}_{\rm a}=(\hat{\rm a}+\hat{\rm a}^\dagger)/\sqrt{2}$, $\hat{Y}_{\rm a}=(\hat{\rm a}-\hat{\rm a}^\dagger)/i\sqrt{2}$, $\hat{X}_c=(\hat{c}+\hat{c}^\dagger)/\sqrt{2}$ and $\hat{Y}_c=(\hat{c}-\hat{c}^\dagger)/i\sqrt{2}$ (and their corresponding input noise quadratures). With the coupling strength ${\mathcal G}=2{\rm g}$, this allows the quantum Langevin equations for the quadratures to be written in a compact matrix form 
\begin{equation}\label{eq_motion}
 \dot{\rm \hat u}(t)=\mathcal{A}{\rm \hat u}(t) +\eta(t),
\end{equation}
where $\hat{u}(t)=\{{\hat X}_{\rm a}, {\hat Y}_{\rm a}, {\hat X}_c, {\hat Y}_c\}^T$ and $\eta(t)=\{\sqrt{2\kappa_{\rm fb}}{\hat X}_{\rm a}^{\rm in},\sqrt{2\kappa_{\rm fb}}{\hat Y}_{\rm a}^{\rm in}, \sqrt{2\kappa_c}{\hat X}_c^{\rm in}, \sqrt{2\kappa_c}{\hat Y}_c^{\rm in}\}^T$ are, respectively, the vector of operators and the vector of noises. The drift matrix $\mathcal{A}$ is given by
\begin{equation}\label{drift}
\mathcal{A} =\left[\begin{array}{*{20}c}
{{-\kappa}_{\rm fb}} & {{\Omega}_{\rm fb}} & {{0}} & {{0}}  \\
{{-\Omega}_{\rm fb}} & {{-\kappa}_{\rm fb}} & {\mathcal G} & {{0}}  \\
{{0}} & {{0}} & {{-\kappa_c}} & {{\Omega_c}}  \\
{\mathcal G} & {{0}} & {{-\Omega_c}} & {{-\kappa_c}}  \\
\end{array}\right],
\end{equation}
if all eigenvalues of the drift matrix $\mathcal{A}$ have negative real parts, the system is stable and reaches a steady-state. The linear dynamics and Gaussian quantum noise terms lead to the system reaching a continuous variable (CV) Gaussian steady-state. This state can be characterized by the $4\times 4$ covariance matrix (CM) $\mathcal{V}$, whose components are given by
\begin{equation}\label{CM}
\mathcal{V}_{ij}= \langle {\rm \hat u}_i (\infty){\rm \hat u}_j (\infty) + {\rm \hat u}_j (\infty){\rm \hat u}_i (\infty)\rangle/2.
\end{equation}
Given that the first moments are zero, covariance matrice (CM) are constrained by the uncertainty relations among canonical operators ($[\hat{u}_i, \hat{u}_j]=i\Omega_{ij}$). This imposes the inequality $\mathcal{V}+i\mathbf{\Omega}/2\succeq0$, where $\mathbf{\Omega}$ is the symplectic matrix \cite{Serafini-BOOK}. The Lyapunov equation characterizes the steady state of the system, and is expressed as
\begin{equation}\label{lyap}
 \mathcal{A}\mathcal{V} + \mathcal{V}\mathcal{A}^T = - \mathscr{D},
\end{equation}
where $\mathscr{D}=diag \{{\rm K_{\rm a}}(2N_{\rm a}+1), {\rm K_{\rm a}}(2N_{\rm a}+1), \kappa_c (2N_{c}+1), \kappa_c (2N_{c}+1) \}$ is the diffusion matrix, with ${\rm K_{\rm a}}=\kappa_{\rm a}(1-\tau^2)|1-\tau {\rm e}^{{\rm i}\theta}|^2$. While Equation (\ref{lyap}) is readily solvable, its full analytical expression is too cumbersome to be presented here. The steady-state entropy production rate ($\Pi_s$) and entropy flux rate ($\phi_s$) are determined to be as presented in the Appendix \cite{Landi,Brunelli1,Brunelli2}
\begin{eqnarray}\label{Pi-s-1}
\Pi_s=-\phi_s&=&{\rm Tr}[\mathcal{A}^{irr} + 2 \mathcal{A}^{irr} \mathscr{D}^{-1} \mathcal{A}^{irr} \mathcal{V}^s ]\nonumber\\
&=&2\kappa_{\rm fb}\bigg[\frac{\kappa_{\rm fb}}{{\rm K}_{\rm a}}\frac{(\mathcal{V}_{11}^s+\mathcal{V}_{22}^s)}{(2N_{\rm a}+1)}-1\bigg]+2\kappa_c\bigg[\frac{\mathcal{V}_{33}^s+\mathcal{V}_{44}^s}{2N_c+1}-1\bigg]\nonumber\\
&=& \mu_{\rm a} + \mu_c,\,\,\,\,\,\,\,\,\,\,\,
\end{eqnarray}
where $\mathcal{A}^{irr}=diag\{-\kappa_{\rm fb},-\kappa_{\rm fb},-\kappa_c,-\kappa_c\}$. When the system is in the equilibrium state: $\mathcal{V}_{11}^s+\mathcal{V}_{22}^s=\frac{{\rm K}_{\rm a}}{\kappa_{\rm fb}}(2N_{\rm a} + 1)$ and $\mathcal{V}_{33}^s+\mathcal{V}_{44}^s = 2N_c + 1$, and hence, $\Pi_s=0$ \cite{Brunelli2,Malouf,Salazar}. We define the first term as contribution ${\rm a}$ to the entropy production rate ($\mu_{\rm a}$), as it depends solely on quantities labeled by ${\rm a}$. Analogously, the second term, dependent on quantities labeled by ${\rm c}$, is defined as contribution ${\rm c}$ ($\mu_{\rm c}$)
\begin{equation}\label{mu_k}
\mu_k=2 \kappa_k \left( \frac{N_{k,\text{s}}+1/2}{N_k+1/2}-1\right),~~~(k={\rm a}, c),
\end{equation}
the terms $N_{{\rm a},\text{s}}$ and $N_{\rm a}$ represent $\langle\hat{{\rm a}}^{\dagger}\hat{{\rm a}}\rangle_\text{s}$ and $\langle\hat{{\rm a}}^{\dagger}\hat{{\rm a}}\rangle_\text{eq}$, respectively, with $\mu_{\rm c}$ defined similarly. Equation~\eqref{mu_k} links the irreversibility generated by the stationary process to the change in each oscillator's excitation level from its equilibrium value. For a noninteracting system, where each oscillator equilibrates with its individual bath, $\Pi_s$ identically vanishes, as demonstrated by Eq.~\eqref{mu_k}.

Mutual information, defined as $\mathcal{I}(\varrho_{{\rm a}:{\rm c}})=S(\varrho_{\rm a})+S(\varrho_{\rm c})-S(\varrho_{{\rm a}{\rm c}})$, quantifies the total correlations shared between two systems~\cite{VlatkoRevEntropy}, using the von Neumann entropy for Gaussian distributions 
\begin{equation}\label{ShanWig}
S(\mathcal{V}_{{\rm a}c})=\frac12 \log (\det \mathcal{V}_{{\rm a}c}).
\end{equation}
According to Ref.~\cite{Adesso}, $S(\mathcal{V}_{{\rm a}{\rm c}})$ coincides (up to an additive constant) with the generalized R\'enyi entropy of order 2 ($S_2(\varrho)=-\log \tr{\varrho^2}$). Mutual information is defined as $\mathcal{I}(\mathcal{V}_{{\rm a}:{\rm c}})=S(\mathcal{V}_{{\rm a}{\rm c}}\vert\vert\mathcal{V}_{\rm a} \oplus \mathcal{V}_{\rm c})$. Moreover, Eq. (\ref{ShanWig}) implies that
\begin{equation}\label{IG}
\I(\mathcal{V}_{{\rm a}:{\rm c}})=-\frac12 \log \left(\frac{\det\mathcal{V}}
{\det\mathcal{V}_{\rm a}\det\mathcal{V}_{c}} \right) \,.
\end{equation}
We quantify bipartite entanglement using the logarithmic negativity $E_N$ \cite{Plenio,Vidal02,Adesso04}, which takes the form
\begin{equation} \label{EN}
E_N = \max[0,-\log(2\nu^-)],
\end{equation}
where, $\nu^{-}$ is the minimum symplectic eigenvalue of the covariance matrix $\mathcal{V}$:$$\nu^-= \frac{1}{\sqrt{2}}\sqrt{\mathcal{L}-(\mathcal{L}^2-4\det\mathcal{V})^{1/2}},$$ with $\mathcal{L}=\det \mathcal{V}_{\rm a}+\det \mathcal{V}_c-2\det \mathcal{V}_{{\rm a}c}$. The condition $\nu_{-} < 1/2$ is the necessary and sufficient criterion for entanglement between the two modes, as it directly corresponds to a positive logarithmic negativity ($E_{N}=-\log(2\nu^-)>0$).

\section{Resultats and discussion}

In this section we will show the importance technique of coherent feedback, that enable one to generate the irreversibility in the equilibrium ($N_{\rm a}=N_c=0$), and also its importance to enhances the irreversibility entropy production.
\begin{figure}[th]
\includegraphics[width=1 \columnwidth]{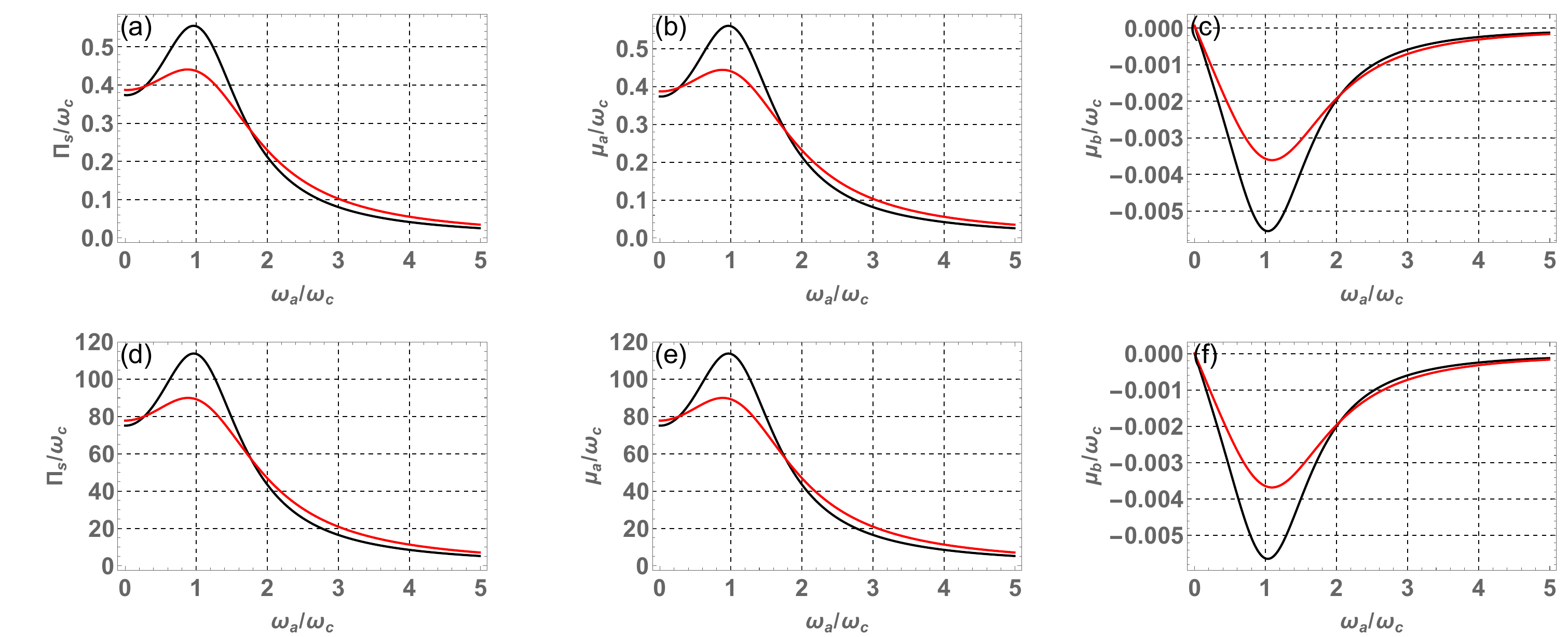} 
\caption{Entropy production rate $\Pi_\text{s}/\omega_c$ and its two contributions $\mu_{\rm a}/\omega_c$ and $\mu_c/\omega_c$ against the ratio of the two frequencies for differents value of $\kappa_c$ with $\kappa_{\rm a}=\kappa_c=0.2\omega_c$ correspond to the Back curves, while $\kappa_{\rm a}=0.2\omega_c$ and $\kappa_c=0.5\omega_c$ correspond to the Red curves. In panels (a), (b), (c) when the reservoirs occupy their ground state $N_{\rm a}=N_c=0$. In panels (d), (e), (f) an imbalance in thermal excitations is considered, where $N_{\rm a}=0$ and $N_c=100$. We use parameters are $G=0.1\omega_c$ $\tau=0.9$ and $\theta=\pi$.}   
\label{figPI1}
\end{figure}
We plot the stationary entropy production rate $\Pi_\text{s}$ and its constituent components $\mu_{\rm a}$ and $\mu_c$ in Fig.~\ref{figPI1} where the reservoirs are in the ground state $N_{\rm a}=N_c=0$ in panels \textbf{(a)}-\textbf{(c)}, versus the frequency $\omega_{\rm a}$ with solid black curves in the plots correspond to identical loss rates $\kappa_{\rm fb}=\kappa_c$ conversely, the solid red curves are for $\kappa_{\rm a}=0.2,\,\kappa_c=0.5$. In Fig.~\ref{figPI1}, $\Pi_\text{s}$, $\mu_{\rm a}$ and $\mu_c$ are increased via $\tau=0.9$. This is compared with the results discussed in Ref.~\cite{Mauro2016}.

Moreover, in panels \textbf{(a)}-\textbf{(c)}, we remark that $\mu_{\rm a}$ is positive, while $\mu_c$ displays a negative dip. However, one can explain the change of sign of $\mu_c$ from positive ($\tau = 0$) to negative ($\tau = 0.9$)
by coherent feedback ($\tau=0.9$), i.e., $\mu_{\rm a}$ and $\mu_c$ are not similar in the ground state. Furthermore, Fig.~\ref{figPI1} shows $\Pi_{\text s}$ approaching zero for $\omega_{\rm a}\gg1$. When the oscillators are far off resonance, they effectively decouple, leading to each thermalizing with its own bath. The notable disparity in magnitude between $\mu_{\rm a}$ and $\mu_{\rm c}$ leads to the overall positivity of $\Pi_{\text s}$.

This can be explained by the emergence of an imbalance in thermal excitation's by increasing the initial thermal occupation of the oscillators with respect to their initial value, through \textit{the coherent feedback loop that re-injects photons into the cavity}. Furthermore, consistent with Eq.~\eqref{mu_k}, a steady negative value of $\mu_{\rm c}$ indicates a reduction in steady excitations ($N_{c,\text{s}}<N_c$) and thus an effective cooling of oscillator $c$, and increases of steady excitations $N_{{\rm a},\text{s}} > N_{\rm a}$ and thus an effective heating of oscillator ${\rm a}$. This directly results from the enforced asymmetry between the two subsystems, a characteristic observable even if $N_{\rm a}=N_c=0$. We observe that $\Pi_{\text s}\approx \mu_{\rm a}$, featuring a distinctive peak at $\omega_{\rm a}=\omega_c$. Moreover, at $\omega_{\rm a}=\omega_c$, we see that $\Pi_\text{s}$, $\mu_{\rm a}$ and $|\mu_c|$ achieve its maximum value for equal loss $\kappa_{\rm a}=\kappa_c$. It is also evident from Fig.~\ref{figPI1}(\textbf{(a)}-\textbf{(c)}) that $\Pi_{\text s}$ generally broadens with increasing losses, $\kappa_{\rm a}$ and $\kappa_{\rm c}$.

If we consider initial thermal occupation in one oscillator, specifically $N_c=100>0$ as shown in Fig.~\ref{figPI1}\textbf{(d)}-\textbf{(f)}, $\Pi_{\text s}$ is found to approximate $\mu_{\rm a}$ and displays a distinctive peak at $\omega_{\rm a}=1$. In parallel, $\mu_{\rm c}$ exhibits a negative dip. This figure show also that coherent feedback lead to enhance the degree of irreversibility, in comparison with the results shown in Ref. \cite{Mauro2016}. A comparison of panels \textbf{(a)} and \textbf{(d)} immediately reveals that employing a coherent feedback loop (under conditions where $N_{\rm a}=N_c=0$) or can be made to grow by introducing an imbalance in the initial oscillator populations $\Pi_\text{s}$. This indicates irreversibility arising from transport, as the coupled oscillators now facilitate a net heat flux between the two baths. Consistent with Eq.~\eqref{mu_k}, the observation of a steady negative $\mu_c$ value signifies a reduction in steady excitations $N_{c,\text{s}} < N_c$ and thus an effective cooling of oscillator $\rm a$.

The peaks in $\mu_{\rm a}$ and corresponding dips in $\mu_{\rm c}$ at $\omega_{\rm a}=1$ (i.e., when the oscillators share identical frequencies). For $\omega_{\rm a}\approx1$ and $G<\omega_{\rm a}$, adopting the rotating frame and applying the rotating-wave approximation yields an interaction Hamiltonian of the form $H_{I}\propto\hat{\rm a}^{\dagger}\hat{\rm c}+\hat{\rm a}\hat{\rm c}^{\dagger}$. The large degree of irreversibility can thus be attributed to the latter being a pure exchange interaction, rendering it optimal for heat transfer~\cite{Asadian2013}. 
 
\begin{figure}[th]
\includegraphics[width=1 \columnwidth]{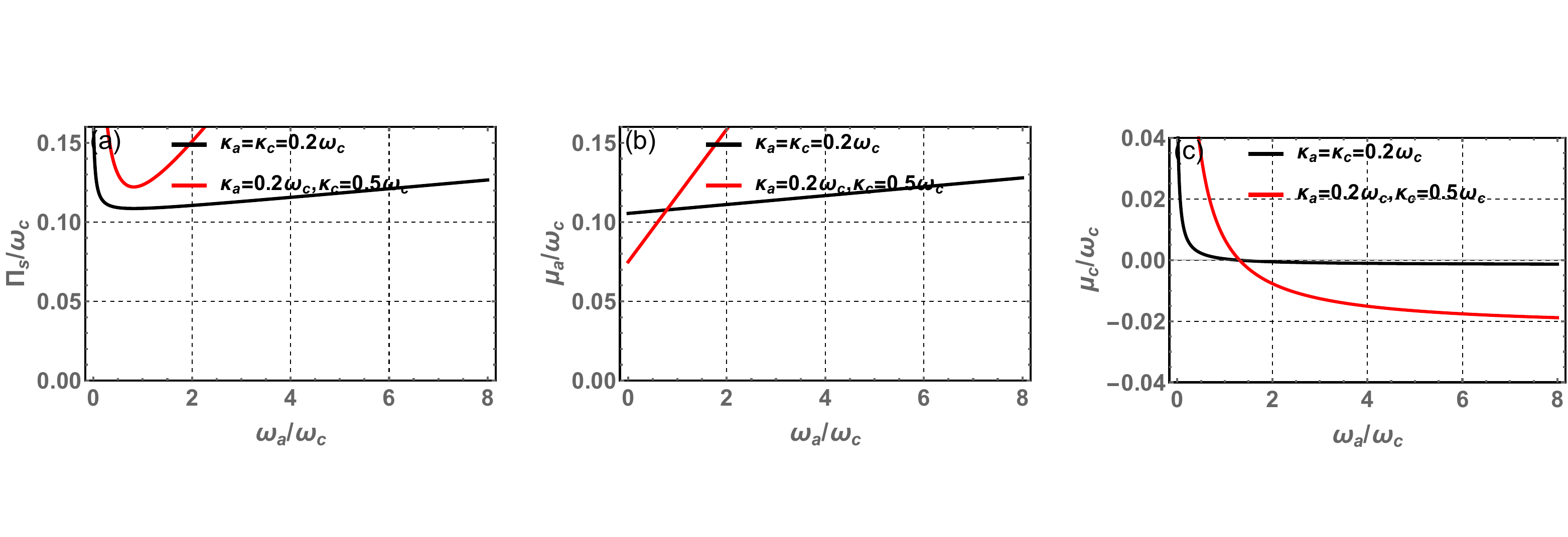} 
\caption{Entropy production rate $\Pi_\text{s}/\omega_c$ (a) and its two contributions $\mu_{\rm a}/\omega_c$ (b) and $\mu_c/\omega_c$ (c) against the ratio $N_{\rm a}/N_c$ for various value of $G$ with $G=0.05\omega_c$ correspond to the Back curves, while $G=0.2\omega_c$ correspond to the Red curves. The oscillators have the same frequency $\omega_{\rm a}=\omega_c$, $\kappa_{\rm a}=0.2\omega_c$ and $\kappa_{\rm c}=0.5\omega_c$, with $\tau=0.1$ and $\theta=\pi$.}   
\label{fig31}
\end{figure}

From Fig.~\ref{fig31}, two distinct behaviors of $\Pi_{\text s}$ are evident: it achieves its minimum at $N_{\rm a}=N_{\rm c}$, while for $N_{\rm c}>N_{\rm a}$, $\Pi_{\text s}$ grows linearly with $N_{\rm c}$, hence proportionally to the population imbalance. Within the regime where $N_c/N_{\rm a}<1$, we observe a sharp increase in the entropy production rate. This rate, however, consistently remains finite, including the case where $N_c=0$. Also, we observe the enhancement of $\Pi_\text{s}$, $\mu_{\rm a}$ and $\mu_c$ via coherent feedback loop. This in comprising with the results realized in Ref. \cite{Mauro2016}. Moreover, we see that $\Pi_\text{s}$ increases with increasing $G$. This can be explained by the relationship between $\mu_{k}~(k={\rm a},c)$ and $G$. Furthermore, for $\tau=0.1$ the $\mu_{\rm a}$ remains positive, even if $N_{\rm a}\ll N_c$. Moreover, $\mu_c$ is negative, around $N_{\rm a}\geq N_c$, as also discussed in Ref. \cite{Mauro2016}. Besides, for $\tau=0.1$ the $\Pi_\text{s}$ remains $>0.1$. This dictates that $\Pi_\text{s}$ will be reduced by the oscillator $c$ and compensate by the oscillator $\rm a$. In the small coupling regime, $\mu_{\rm a}$ and $\mu_{\rm c}$ effectively swap their dependencies on the thermal excitations $N_{\rm a}$ and $N_{\rm c}$. The clear influence of the imbalance between $N_{\rm a}$ and $N_c$ on $\mu_{\rm a}$ and $\mu_{\rm c}$ is evident from this figure, consequently determining which oscillator experiences local entropy reduction and which performs compensation.

\begin{figure}[th]
\includegraphics[width=1 \columnwidth]{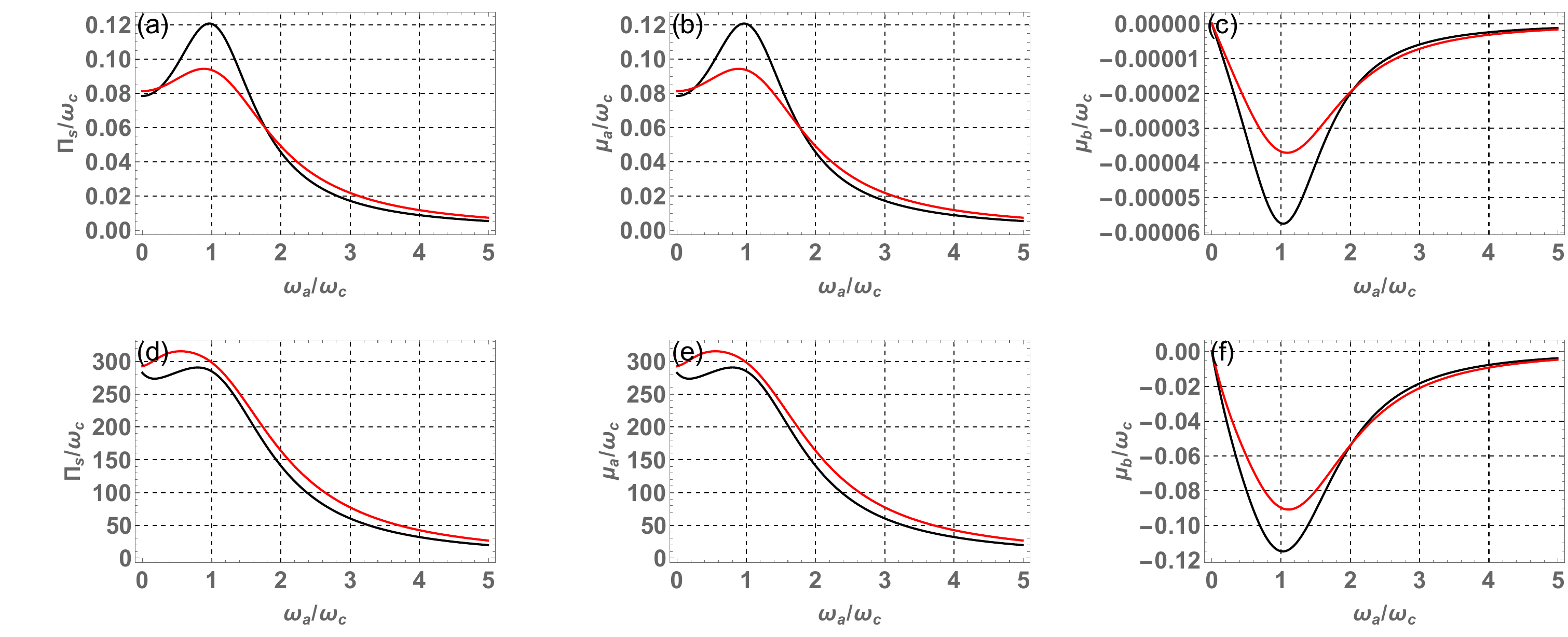}
\caption{Entropy production rate $\Pi_\text{s}/\omega_c$ and its two contributions $\mu_{\rm a}/\omega_c$ and $\mu_c/\omega_c$ against the ratio of the two frequencies for differents value of $\kappa_c$ with $\kappa_{\rm a}=\kappa_c=0.2\omega_c$ correspond to the Back curves, while $\kappa_{\rm a}=0.2\omega_c$ and $\kappa_c=0.5\omega_c$ correspond to the Red curves. In panels (a)-(c) we use $G=10^{-2}\omega_c$ but in panels (d)-(f) we use $G=0.6\omega_c$. Other parameters are $N_{\rm a}=0$ and $N_c=10$, $\tau=0.9$ and $\theta=\pi$.}   
\label{figPI2}
\end{figure}
We explore entropy production rate $\Pi_\text{s}/\omega_c$ and its two contributions $\mu_\text{a}/\omega_c$ and $\mu_c/\omega_c$, as in Fig.~\ref{figPI1} with $N_\text{a}=0$ and $N_c=10$, when coherent feedback is applied. While Fig.~\ref{figPI2}\textbf{(a)-(c)} ($G=10^{-2}\omega_c$) and Fig.~\ref{figPI2}\textbf{(d)-(f)} ($G=0.6\omega_c$) show $\Pi_{\text s}/\omega_c$ having the same shape, a striking difference is its magnitude, which decreases by two orders. In Fig.~\ref{figPI2} \textbf{(d)-(f)}, we observe that the $\Pi_\text{s}/\omega_c$ for $\kappa_{\rm a}=\kappa_{\rm a}=0.2\omega_c$ is bounded by the one of $\kappa_{\rm a}=0.2\omega_c$ and $\kappa_{\rm a}=0.5\omega_c$. For $\tau=0.9$, we remark the enhancement of irreversibility entropy production $\Pi_\text{s}/\omega_c$ and its two contributions $\mu_\text{a}/\omega_c$ and $\mu_c/\omega_c$ with increasing $G$, with respect to the results discussed in Ref. \cite{Mauro2016}. Additionally, $\Pi_{\text s}$ shows a second peak around $\omega_{\rm a}/\omega_{\rm c}=1$, consistent with hybridization effects.
\begin{figure}[th]
\includegraphics[width=1 \columnwidth]{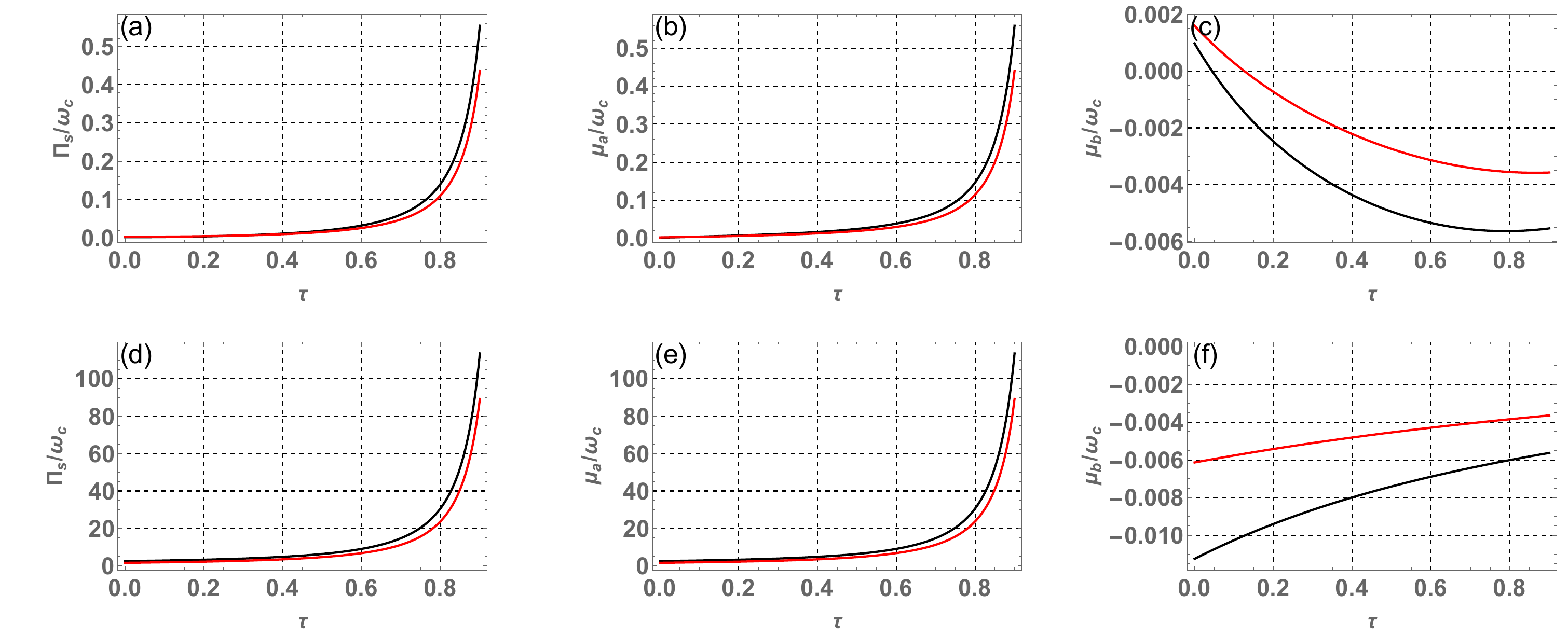}  
\caption{Entropy production rate $\Pi_\text{s}/\omega_c$ and its two contributions $\mu_{\rm a}/\omega_c$ and $\mu_c/\omega_c$ against the reflectivity parameter $\tau$ for various value of $\kappa_c$ with $\kappa_{\rm a}=\kappa_c=0.2\omega_c$ correspond to the Back curves, while $\kappa_{\rm a}=0.2\omega_c$ and $\kappa_c=0.5\omega_c$ correspond to the Red curves. In panels (a), (b), (c) the reservoirs are in the ground state $N_{\rm a}=N_c=0$, while in panels (d), (e), (f) we considering an imbalance in thermal excitations $N_{\rm a}=0$ and $N_c=100$, with $\theta=\pi$ and $G=0.1\omega_c$.}   
\label{figtau}
\end{figure}

We present in Fig. \ref{figtau}\textbf{(a-f)} the evolution of entropy production $\Pi_\text{s}/\omega_c$ and its components $\mu_{\rm a}/\omega_c$ and $\mu_c/\omega_c$ versus the reflectivity parameter $\tau$ for various values of $\kappa_{{\rm a}(c)}$ and $N_{{\rm a}(c)}$. We notice the enhancement of $\Pi_\text{s}/\omega_c$, $\mu_{\rm a}/\omega_c$ and $\mu_c/\omega_c$ by increasing $\tau$. Moreover, the coherent feedback contribute on a steady negative value of $\mu_c$, i.e., an effective cooling of oscillator $c$, and on an effective heating of oscillator $\rm a$, because $\mu_{\rm a}$ increases. Also, an imbalance in thermal excitation’s increases the entropy production. However, the entropy production $\Pi_\text{s}/\omega_c$ increase for $\kappa_{\rm a}=\kappa_c=0.2\omega_c$.


\begin{figure}[th]
\includegraphics[width=1 \columnwidth]{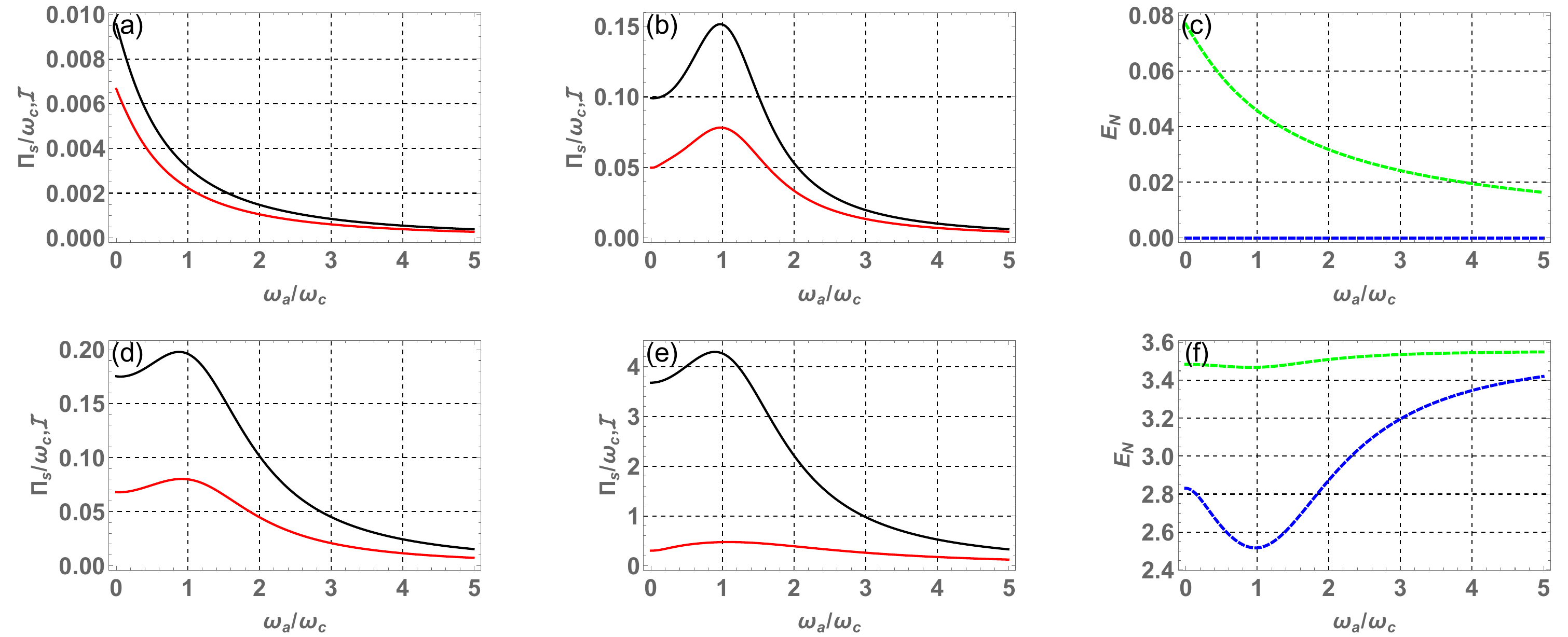}  
\caption{Comparison of the steady-state entropy production rate $\Pi_\text{s}/\omega_c$ (Black curves), the mutual information $\mathcal{I}$ (Red curves) and entanglement (Green and Blue curves) versus the ratio of the two frequencies $\omega_{\rm a}/\omega_c$ with $G=0.1\omega_c$. In Panels (a), (d), (c) and (f) (Green and Blue curve) corresponds to $N_{\rm a}=N_c=0$, while in panels (b), (e), (c) and (f) (Blue curve) corresponds to $N_{\rm a}=0$, $N_c=10$. With $\tau = 0$ in Panels (a), (d), (c) and $\tau = 0.85$ in Panels (b), (e), (f). With $\theta=\pi$.}   
\label{figPI}
\end{figure}

The striking similarity between $\Pi_\text{s}$ and $\mathcal{I}$, shown in Fig. \ref{figPI}\textbf{(a), (b), (d)} and \textbf{(e)}, reveals that $\mathcal{I}$ is a one-to-one function of $\Pi_\text{s}$, despite their different functional forms. This result provides the first indication that the irreversibility arising from the stationary process is closely linked to the total correlations shared between the two modes. Decoupled oscillators (uncoupled or far-off-resonance) reach separate thermal equilibrium, leading to vanishing $\Pi_\text{s}$ and $\mathcal{I}$. it can clearly seen that in the presence of coherent feedback ($\tau=0.85$), $\Pi_\text{s}$, $\mathcal{I}$ and entanglement enhance. Besides, the increases of $\tau$ leads to enhancement of the region around the resonance, i.e., the two oscillators go far of equilibrium. Furthermore, panels (d) and (e) illustrate the enhancement of both $\Pi_\text{s}$ and $\mathcal{I}$ relative to the results shown in panels (a) and (b), respectively. In panel (c), the entanglement is vanishing for $N_{\rm a}=0$, $N_c=10$. However, $E_N$ is positive for $N_{\rm a}=N_c=0$. This means that the entanglement is fragile under thermal effect than both $\Pi_\text{s}$ and $\mathcal{I}$. it can be seen that when $\tau=0.86$, the entanglement is generated by coherent feedback when $N_{\rm a}=0$, $N_c=10$. 

\section{Application : cavity optomechanics}

\begin{figure}
\centering
 \includegraphics[width=12cm,height=6cm,angle=0]{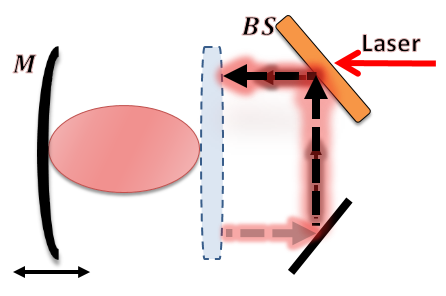}
 \caption{Schematic of Fabry-P\'erot cavity with coherent feedback. An input laser is coupled into the cavity through an asymmetric beam splitter (BS), which also routes a portion of the cavity's output field back into the cavity via a reflecting mirror.}
 \label{Fig2} 
\end{figure}

Characterization of the irreversibility in a cavity optomechanical system is performed in this section, utilizing our previously established results. Configured as a single-mode Fabry-P\'erot cavity, this system includes a perfectly reflecting movable mirror (Fig. \ref{Fig2}). Pumping the length-${\rm L}$ cavity through the fixed mirror is a monochromatic laser field, characterized by frequency $\omega_0$ and strength $\mathcal{\rm E}=\sqrt{\frac{2 {\rm P} \kappa_{\rm a}}{\hbar \omega_0}}$ (${\rm P}$ = laser power, $\kappa$ = cavity decay rate). The optomechanical coupling between the cavity and the mechanical oscillator, with frequency $\omega_{\rm m}$, is enabled by the radiation pressure. In the rotating frame of the external pump, the system's Hamiltonian ($\hbar = 1$) is given by
\begin{equation}\label{HamOpto}
\mathsf{H}=\hbar\Delta_0\hat{\mathrm{a}}^{\dagger}\hat{\rm a}+\hbar \omega_m \hat{\rm c}^{\dagger}\hat{\rm c} - \hbar \mathsf{g}_0 \hat{\rm a}^{\dagger}\hat{\rm a} (\hat{\rm c}^{\dagger} + \hat{\rm c})
+i \hbar\mathcal{\rm E}\xi(\hat{\rm a}^{\dagger}-\hat{\rm a}) \, ,
\end{equation}
The first two terms in the Hamiltonian $\mathsf{H}$ describe the energies of the individual cavity field and mechanical mode. Let $\Delta_0=\omega_{\rm a}-\omega_0$ be the cavity detuning, and $\mathsf{g}_0$ be the single-photon optomechanical coupling strength, which quantifies the interaction between the cavity and the mechanical oscillator. The Hamiltonian's third term captures the radiation-pressure interaction, while the final term depicts the cavity driving in the presence of coherent feedback. For the beam splitter, $\xi$ represents the real amplitude transmission parameter and $\tau$ the real amplitude reflectivity parameter. Both are real, positive, and satisfy $\xi^2 + \tau^2 = 1$ \cite{Mandel}. When coherent feedback of the cavity output is absent, the parameters are $\tau=0$ and $\xi=1$. We can obtain the following nonlinear quantum Langevin equations through the inclusion of dissipation arising from system-bath interactions. By including dissipation arising from system-bath interactions, we obtain the following nonlinear quantum Langevin equations

\begin{subequations}\label{Langevin_1}
\begin{eqnarray}
\dot{\hat{{\rm a}}}&=&-(i \Delta_{\rm fb} + \kappa_{\rm fb}) \hat{{\rm a}}+ i \mathsf{g} \hat{{\rm a}}(\hat{{\rm c}}+\hat{{\rm c}}^\dagger) + \mathcal{{\rm E}}+ \sqrt{2\kappa} {\rm a}_{\rm fb}^{in},\\
 \dot{\hat{{\rm c}}}&=&-(i \omega_{\rm c} + \gamma_{\rm c})\hat{{\rm c}} +i \mathsf{g} \hat{{\rm a}}\hat{{\rm a}}^\dagger+\sqrt{2\gamma} {\rm c}^{in},\nonumber\\
\end{eqnarray}
\end{subequations}
${\rm a}_{\rm fb}^{in}$ and ${\rm c}^{in}$ are input-noise operators, $\gamma_{\rm c}$ is the mechanical oscillator damping rate, and $\kappa_{\rm fb}$ is the cavity-field damping rate via feedback. When the driving laser amplitude is strong, both modes exhibit large steady-state amplitudes. We express each operator as the sum of its steady-state mean and a fluctuation: $\hat{{\rm a}}=\hat{{\rm a}}_s+\delta \hat{{\rm a}}$ and $\hat{{\rm c}}=\hat{{\rm c}}_s+\delta \hat{{\rm c}}$. The steady-state mean values are expressed as
$$\hat{{\rm a}}_s=\frac{\mathcal{{\rm E}}}{i \Delta'_{\rm fb} + \kappa_{\rm fb}},\quad \hat{{\rm c}}_s=\frac{i\mathsf{g}_0|\hat{{\rm a}}_s|^2}{i \omega_{\rm c} + \gamma_{\rm c}},$$
where $\Delta'_{\rm fb}=\Delta_{\rm fb}-\mathsf{g}_0(\hat{c}_s+\hat{{\rm c}}^*_s)$ defines the effective cavity detuning.
The linearized quantum Langevin equations governing the fluctuation operators are
\begin{subequations}\label{17}
\begin{eqnarray}
\dot{\delta\hat{\rm a}}&=&-(i \Delta_{\rm fb} + \kappa_{\rm fb}) \delta\hat{\rm a}+ i \mathsf{g} (\delta\hat{\rm c}+\delta\hat{\rm c}^\dagger) + \sqrt{2\kappa} \delta {\rm a}_{\rm fb}^{in},\\
 \dot{\delta\hat{\rm c}}&=&-(i \omega_{\rm c} + \gamma_{\rm c})\delta\hat{\rm c} +i \mathsf{g} (\delta\hat{\rm a} + \delta\hat{\rm a}^\dagger) +\sqrt{2\gamma} \delta {\rm c}^{in},\nonumber\\
\end{eqnarray}
\end{subequations}
where $\mathsf{g}$ is defined as $\mathsf{g}_0|\hat{\rm a}_s|$, representing the light-enhanced optomechanical coupling. From comparing Eq. (\ref{17}) and Eq. (\ref{La}), it is evident that the harmonic oscillator in contact with the thermal bath corresponds to the cavity-field fluctuation (with effective frequency $\Omega_{\rm a}=\Delta$ and damping rate $\kappa_{\rm fb}$), while the harmonic oscillator in contact with the thermal bath corresponds to the mechanical mode fluctuation (with frequency $\omega_{c}=\omega_m$ and damping rate $\gamma_{c}=\kappa_c$).

\begin{figure}[th]
\includegraphics[width=1 \columnwidth]{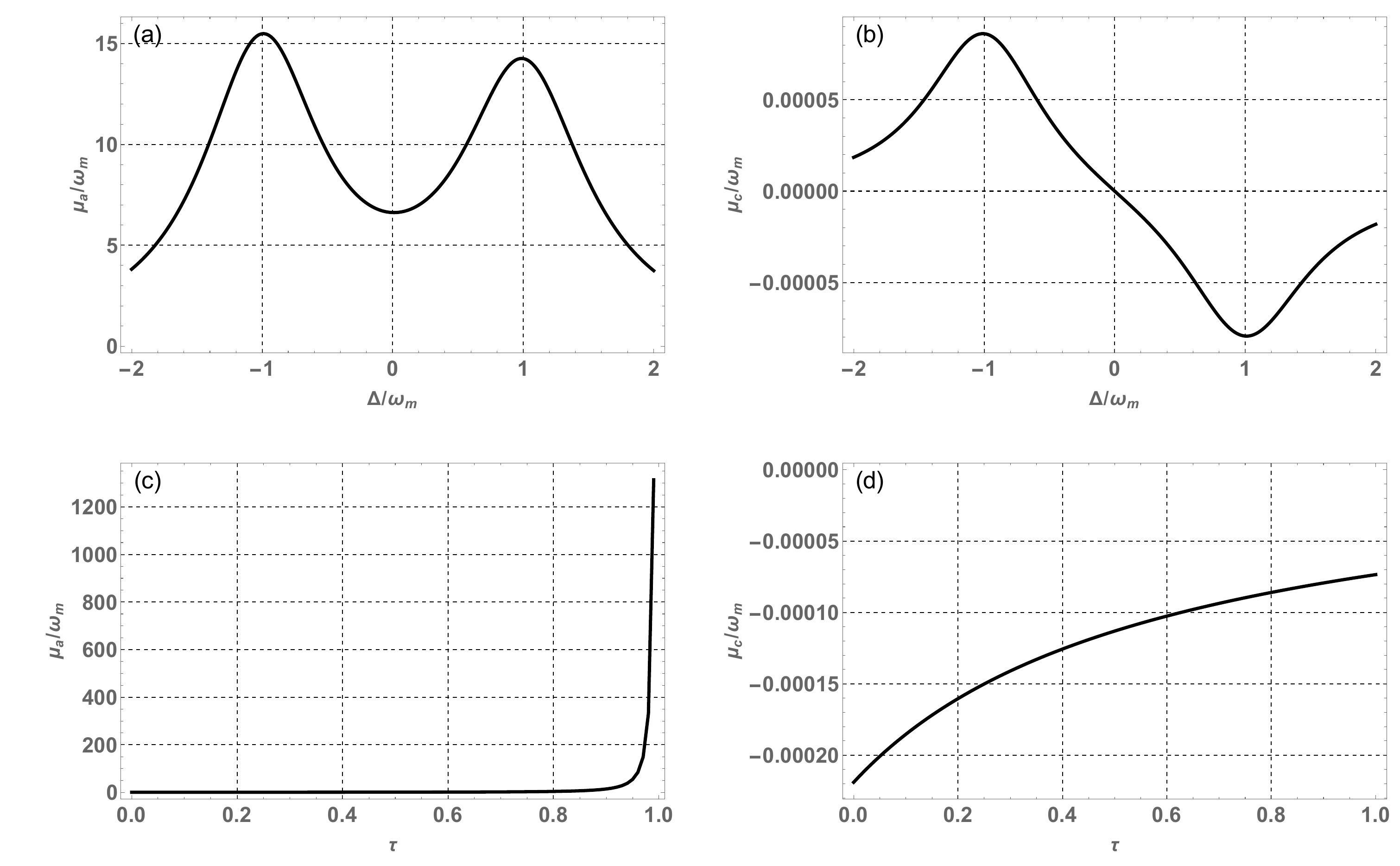} 
\caption{Plot of the optical and mechanical contributions to the entropy production $\mu_\text{a}/\omega_m$ and $\mu_\text{c}/\omega_m$ as a function of normalized photon detuning $\Delta/\omega_m$ with $\tau=0.9$ (a)-(b), and of the reflectivity parameter $\tau$ with $\Delta=\omega_m$ (c)-(d). Other parameters $\kappa_{\rm a}=0.2\omega_m$, $\kappa_c=10^{-3}\omega_m$, ${\rm g} = 0.005\omega_m$, $N_{\rm a}=0$ and $N_c=10^3$. With $\theta=\pi$.}   
\label{figOMs}
\end{figure}

The behavior of the optical ($\mu_\text{a}$) and mechanical ($\mu_\text{c}$) contributions to the entropy production against the normalized detuning is displayed in Fig.~\ref{figOMs}\textbf{(a)-(b)} for the case of coherent feedback. These plots reveal that for small coupling, the system is stable even in the blue-detuned region ($\Delta<0$). Large detuning ($|\Delta| \gg 1$) effectively decouples the two oscillators, causing each to equilibrate with its bath, which results in vanishing $\Pi_\text{s}$.We observe that the contributions $\mu_\text{a}$ and $\mu_\text{c}$ to the entropy production peak at $\Delta=\pm \omega_m$. The peaks in the rate of entropy production at $\Delta>0$ and $\Delta<0$ demonstrate that the cooling and heating processes associated with these detuning regimes behave distinctly. This is shown by the entropy production rate's behavior, with clear peaks in both the optical ($\mu_\text{a}$) and mechanical ($\mu_\text{c}$) contributions at the two mechanical sidebands. For $\Delta\approx \omega_m$, the dominant beam splitter interaction $\hat{H}_I \propto \delta \hat{\rm a}^\dagger \delta \hat{\rm c} + \delta \hat{\rm a} \delta \hat{\rm c}^\dagger$ causes enhanced heat transport and a necessary increase in entropy. In the amplification regime, the peak is observed to be even more pronounced, as seen in Fig.~\ref{figOMs}\textbf{(a)}. In the regime $\Delta\approx -\omega_m$, the dominant two-mode squeezing interaction $\hat{H}_I \propto \delta \hat{\rm a}^\dagger \delta \hat{c}^\dagger+\delta \hat{\rm a} \delta \hat{c}$ results in exponential growth of oscillator energies and induces strong correlations between the two modes, yielding EPR-like entanglement in the infinite energy limit. Consequently, our analysis indicates that the entropy production must increase accordingly. Conversely, the sign change of $\mu_b$ in Fig.~\ref{figOMs}\textbf{(b)} clearly indicates the mechanical resonator's heating/cooling. The behavior of $\Pi_{\text s}$ and its contributions thus provides a full thermodynamical account of both the amplification and cooling regimes.

The optical ($\mu_\text{a}$) and mechanical ($\mu_\text{c}$) contributions to the entropy production are plotted as a function of the reflective parameter $\tau$ in Fig.~\ref{figOMs}\textbf{(c)-(d)}. These figures reveal a significant enhancement of entropy production through coherent feedback. By re-injecting photons into the cavity, coherent feedback enhances the interaction between the optical and mechanical modes.

\begin{figure}[th]
\includegraphics[width=1 \columnwidth]{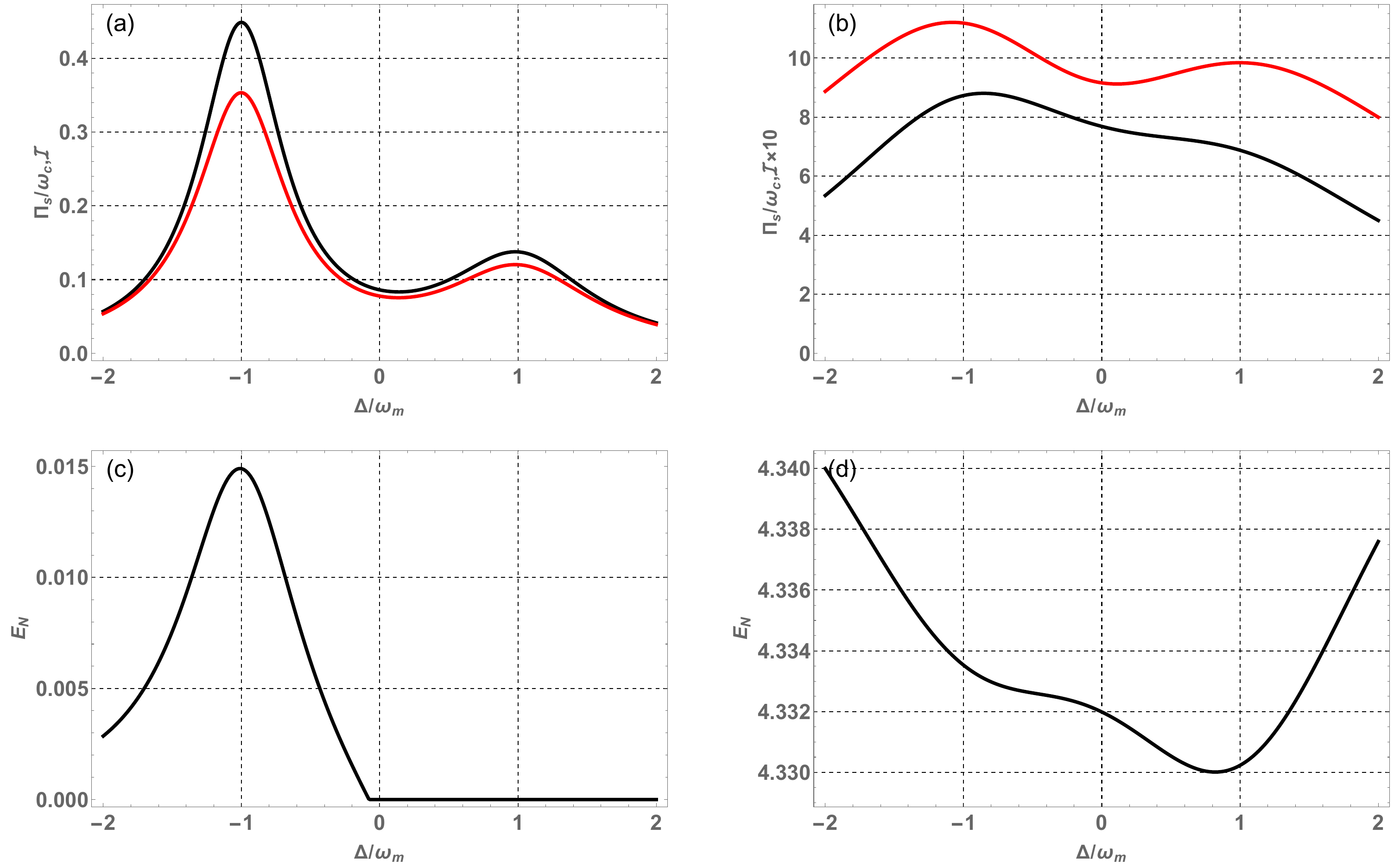} 
\caption{Plot of the optical and mechanical contributions to the entropy production $\mu_\text{a}/\omega_m$ and $\mu_\text{c}/\omega_m$ as a function of normalized photon detuning $\Delta/\omega_m$ with $\tau=0$ in (a) and (c); and $\tau=0.9$ in (b) and (d). Other parameters $\kappa_{\rm a}=0.5\omega_m$, $\kappa_c=10^{-2}\omega_m$, ${\rm g} = 0.05\omega_m$, $G=2{\rm g}$, $N_{\rm a}=0$ and $N_c=10$. With $\theta=\pi$.}   
\label{EPIOMs}
\end{figure}

Fig.~\ref{EPIOMs}\textbf{(a)} and \textbf{(b)} present a comparison between the entropy production rate $\Pi_{\text s}$ and the mutual information $\mathcal{I}$, which quantifies the correlations established by the optomechanical interaction. It can be seen that for small thermal phonon excitation ($N_c =10$), Fig.~\ref{EPIOMs}\textbf{(a)} and \textbf{(b)} reveal that $\Pi_{\text s}$ and $\mathcal{I}$ are nearly identical across the entire range of detunings $\Delta$. This indicates that the degree of irreversibility induced by the stationary process is directly related to the amount of shared correlations. In addition, $\Pi_{\text s}$ and $\mathcal{I}$ exhibit strong peaks around $\Delta = -\omega_m$. This is attributable to the strong quantum correlations (entanglement) established between the optical and mechanical modes, evident from Fig.~\ref{EPIOMs}\textbf{(a)}. Increasing the reflective parameter $\tau$ leads to an increase in the entropy production rate $\Pi_{\text s}$, mutual information $\mathcal{I}$, and entanglement $E_N$, as demonstrated in Fig.~\ref{EPIOMs}\textbf{(b)} and \textbf{(d)}. 

\section{Conclusion}

In summary, we have investigated the enhancement of the entropy production rate through the coherent feedback loop in a cavity optomechanical system where an optical cavity mode is coupled with movable mirror via radiation pressure. We have shown that in the small-coupling limit, the entropy production rate is proportional to the quantum mutual information. We use for application the optomechanical system of Fabry-P\'erot type cavity, and show that the picks of the entropy production corresponding of the heating/cooling of movable mirror are improved. Therefore, we conclude that irreversibility and quantum correlations are not independent and must be analyzed jointly and could be enhanced via feedback controle.

\section*{APPENDIX : Stationary entropy production rate}

Explicit expressions for the entropy production rate $\Pi(t)$ and the entropy flux $\Phi(t)$ from Eq.~\eqref{RateEq} are provided in this Appendix, following the approach of Refs. \cite{Landi,Brunelli1,Brunelli2}. The system's dynamics are described by the quantum Langevin equations for the vector $\hat u=(\hat X_{\rm a}, \hat Y_{\rm a}, \hat X_c, \hat Y_c)^T$. 
The relationship between the Wigner entropy and the covariance matrix ($\mathcal{V}$) for a Gaussian system is given by:
\begin{equation}\label{W-form}
\mathscr{W}(u)=\frac{1}{\pi^n\sqrt{det\mathcal{V}}}e^{-\frac{1}{2}u^T\mathcal{V}^{-1}u},
\end{equation}
This property of being always positive allows us to identify the Wigner function as a quasi-probability distribution in phase space ($n$ is the number of bosonic modes). It provides a fully equivalent description of the density matrix. According to Ref. \cite{Adesso}, the Wigner entropy is related to the R\'enyi-2 entropy and satisfies the strong subadditivity inequality. In recent years, the link between general R\'enyi-$\alpha$ entropies and the thermodynamic properties of quantum systems has also been a subject of interest \cite{Wei,Brandao}. An equivalent description of the system's dynamics is given by the Fokker-Planck equation for the Wigner function $\mathcal{W}(u,t)$ (Eq.~\eqref{W-form})
\begin{equation}\label{WJ} 
\frac{{\rm d} \mathscr{W}}{{\rm d} t} + \frac{{\rm d} {\rm J}(u,t)}{{\rm d}u}=0  \, ,
\end{equation}  
where $\hat u=(\hat X_{\rm a}, \hat Y_{\rm a}, \hat X_c, \hat Y_c)^T$ is a point in the phase space. The probability current vector is then defined as 
\begin{equation}\label{J}
{\rm J}(u,t)=\mathcal{A} u \mathscr{W}(u,t)-\frac{\mathscr{D}}{2} \frac{{\rm d} \mathscr{W}(u,t)}{{\rm d} u} \, .
\end{equation}
the drift matrix $\mathcal{A}$ and the diffusion matrix $\mathscr{W}$ defined in Eq.~\eqref{lyap}. Using the time-reversal operator ${\rm E}=\text{diag}(1,-1,1,-1)$, the dynamical variables are split according to their symmetry. Accordingly, the drift matrix $\mathcal{A}$ is decomposed into an irreversible component $\mathcal{\Airr}$ and a reversible component $\mathcal{\Arev}$ ($\mathcal{A}=\mathcal{\Arev}+\mathcal{\Airr}$), where the irreversible component $\mathcal{\Airr}$ is odd under time reversal~\cite{Spinney12}. Using the time-reversal operator, the irreversible component $\mathcal{\Airr}$ and the reversible component $\mathcal{\Arev}$ are constructed via $\mathcal{\Airr}=[\mathcal{A}+{\rm E}\mathcal{A}{\rm E}^T]/2$ and $\mathcal{\Arev}=[\mathcal{A}-{\rm E}\mathcal{A}{\rm E}^T]/2$. Their explicit forms are as follows
\begin{equation}\label{Airr}
\mathcal{\Arev} =
\begin{pmatrix}
0 & \Omega_{\rm fb}& 0 & 0 \\
-\Omega_{\rm fb}& 0& G & 0 \\
0 & 0& 0 & \Omega_c \\
G & 0& -\Omega_c & 0 \\
\end{pmatrix}\, \quad \text{and}\quad \mathcal{\Airr}=\text{diag}\left(-\kappa_{\rm fb},-\kappa_{\rm fb},-\kappa_c,-\kappa_c\right) \, .
\end{equation}
While the diffusion matrix $\mathscr{D}$ is purely irreversible ($\mathscr{D}^{\text{rev}}=0, \text{ so } \mathscr{D} \equiv \mathscr{D}^{\text{irr}}$), this separation induces a similar splitting in the probability current ${\rm J}(u,t)={\rm J}_\text{rev}(u,t)+{\rm J}_\text{irr}(u,t)$, where:
\begin{equation}\label{J}
\rm{\Jrev(u,t)}= \mathcal{\Arev} u\mathscr{W}(u,t)\, , \quad \text{and} \quad \rm \Jirr(u,t)= \mathcal{\Airr} u\mathscr{W}(u,t)-\frac{\mathscr{D}}{2} \frac{{\rm d} \mathscr{W}(u,t)}{{\rm d} u} \, .
\end{equation}
The reversible part of the probability current is divergence-less ($\nabla_u \cdot {\rm J}_\text{rev}(u,t)=0$), which can be seen from the form of Eq.~\eqref{J}. Substituting Eq.~\eqref{J} into the expression for the entropy rate yields \cite{Mauro2016}
\begin{equation} \label{dS}
\frac{\text{d}\rm S}{\text{d} t}=-\frac14 \int \bigg(\frac{{\rm d} \mathscr{W}(u,t)}{{\rm d} t}\bigg) \log \mathscr{W}(u,t) \text{d}u \, .
\end{equation}
The entropy rate can be expressed as, using Eqs.~\eqref{WJ} and \eqref{dS} and integrating by parts
\begin{equation}
\frac{\text{d}\rm S}{\text{d} t}=-\frac14 \int \text{d}u \{\rm \Jirr(u,t)\}^T\frac{1}{\mathscr{W}(u,t)}\frac{{\rm d} \mathscr{W}(u,t)}{{\rm d} u} \, .
\end{equation}
Substitution of Eq.~\eqref{J} yields the splitting shown in Eq.~\eqref{RateEq}
between entropy flux
\begin{equation}\label{Phi}
\Phi(t)=-\frac14 \int \text{d}u \, 2{\rm \Jirr}(u,t)^T \mathscr{D}^{-1}{\rm \Jirr} u \, ,
\end{equation}
and entropy production rate 
\begin{equation}\label{PiRate}
\Pi(t)=\frac14 \int  \text{d}u \frac{2}{\mathscr{W}(u,t)}{\rm \Jirr}(u,t)^T \mathscr{D}^{-1}{\rm \Jirr}(u,t) \, .
\end{equation}
The expression in Eq. \ref{PiRate} is identified with the entropy production rate due to its non-negativity, since Eq.~\eqref{PiRate} represents the integral of a quadratic form. For a Gaussian state, the relation $\{\frac{{\rm d} \mathscr{W}}{{\rm d} u}\}\mathcal{V}(u,t)=-\mathscr{W}_{\mathcal{V}}(u,t)\mathcal{V}^{-1}(t)u$ holds, which yields the irreversible component of the probability current as ${\rm J}^{\text{irr}}(u,t)=\mathscr{W}_{\mathcal{V}}(u,t)[\mathcal{\Airr} +\mathscr{D}\mathcal{V}^{-1}(t)]u$. In the stationary state, the entropy production rate equals the negative of the entropy flux ($\Pi_\text{s}=-\Phi_\text{s}$), whose expression is \cite{Mauro2016}
\begin{equation}\label{Pi_s}
\Pi_\text{s}=\tr\{\mathcal{\Airr}\}+2\tr\{(\mathcal{\Airr})^T\mathscr{D}^{-1}\mathcal{\Airr}\mathcal{V}_\text{s}\}\,.
\end{equation}
$\Pi_\text{s}$ can be given in matrix form as
\begin{equation}\label{Pi_s}
\Pi_\text{s}= 2 \kappa_{\rm a} \left( \frac{[\mathcal{V}_\text{s}]_{11}+[\mathcal{V}_\text{s}]_{22}}{2N_{\rm a} +1}-1\right)   + 
2 \kappa_c \left( \frac{[\mathcal{V}]_{33}+[\mathcal{V}_\text{s}]_{44}}{2N_{c} +1}-1\right). \,
\end{equation}

\end{document}